\documentstyle [12pt]{article}
\title{
\hspace{3.0truein}{\small IFT-490-UNC}\\
\vspace{0.2truein}
{
Phase Transition in $(2+1)d$ Quantum Gravity}
}
\author{Wei Chen\footnotemark[1]\\
Department of Physics\\
University of North Carolina\\
Chapel Hill, NC 27599-3255}
\date{PACS numbers: 11.30.G, 05.70.F}
\footnotetext{$^*$ chen@physics.unc.edu}

\begin{document}
\maketitle
\vspace{0.2truein}
\begin{abstract}
$(2+1)$ dimensional gravity is equivalent to
an exactly soluble non-Abelian Chern-Simons gauge
field theory \cite{W3}. Regarding this as the
topological phase of quantum gravity in $(2+1)d$,
we suggest a topological symmetry breaking by
introducing a mass term for the gauge fields,
which carries a space-time
metric and local dynamical degrees of freedom.
We consider the finite temperature behavior of
the symmetry broken phase, and claim a restoration of the
topological invariance at some critical temperature.
The phase transition is shown of the zeroth order.

\end{abstract}

\newpage
\baselineskip=22.0truept

One physical motivation for studying topological
quantum field theories is that they may describe
a phase of unbroken diffeomorphism
invariance in quantum gravity
\cite{W3}\cite{Ts}\cite{DJT}\cite{Ho}.
This relates closely to the argument that
a fundamental theory of quantum gravity with high
symmetry will not involve a spacetime metric \cite{Pe}.
And it is rather appealing to imagine
that a metric and local
dynamics might arise as some form of
symmetry breaking, and accordingly a topological phase
transition might happen in the early stage of Universe
when it was hot and small in size.

While correct description for quantum
gravity in the macroscopic
physical reality is still to be
discovered, life in $(2+1)d$ seems
easier. As shown by E. Witten \cite{W3},
$(2+1)d$ general relativity
is equivalent to a topological theory
- the Chern-Simons gauge theory
for the Poincare group.
And the latter, as a quantum
field theory, is exactly soluble \cite{W2}.
With no much thought at the moment about
dynamical mechanism for topological
symmetry breaking even in $(2+1)d$, in this work
we like to introduce
into the Chern-Simons action a mass term
that involves a spacetime metric and,
as to be seen, carries local dynamical
degrees of freedom. Therefore it apparently
breaks diffeomorphism and gauge
invariance. This new theory can be regarded
as a symmetry broken phase of
$(2+1)d$ quantum gravity. Then we are interested
in knowing what would
happen if one heats up the system.
While breaking and restoration of various
symmetries at high energy and/or finite
temperature have been known
for a long time in quantum field theories
\cite{CE}\cite{L}\cite{DJ}\cite{W},
here we see an example of phase transition
concerning topological and gauge symmetries.
The transition, classified as zeroth order,
is thus far as we know the most discontinuous -
even the free energy density has a step at the critical
temperature.

The equivalence between $(2+1)d$ general relativity and a
Chern-Simons gauge theory has something to do with
the following analog: the $(2+1)d$ spacetime is flat as
the Ricci tensor vanishes, $R_{\mu\nu}=0$;
while the stationary points of Chern-Simons action are the
flat connections, for which the curvature vanishes,
$F_{\mu\nu} = 0$.
Precisely, by interpreting the vierbein
and spin connection
as gauge fields $A_\mu^a$, the $(2+1)d$
Einstein action
is mapped to the Chern-Simons action \cite{W3}
\begin{equation}
I_{CS} = \frac{1}{8\pi\alpha}\int_\Omega
\epsilon_{\mu\nu\lambda}
\left(A^a_\mu\partial_\nu A^a_\lambda
+ \frac{1}{3}\epsilon^{abc}A^a_\mu A^b_\nu A^c_\lambda\right)\;,
\label{Scs}
\end{equation}
where the space-time manifold $\Omega$ has Lorentzian signature.
With structure
constant $\epsilon^{abc}$,
the gauge group
is so chosen that it is exactly the $(2+1)d$
Poincare group   $ISO(2,1)$,
in which a non-degenerate metric on the Lie algebra exists.
The variation of $A^a_\mu$ under a non-Abelian
gauge transformation should be
$\delta A_\mu^a = D_\mu \tau^a,$
with $D_\mu\tau^a = \partial_\mu\tau^a
+\epsilon^{abc}A^b_\mu\tau^c$.
Reference no spacetime metric, (\ref{Scs}) is
invariant under diffeomorphism transformations as well, and
thus is known as a topological theory.
However, subject to the flat connection condition,
diffeomorphism transformations
completely fall into gauge transformations.
To see this, pick up a vector $V^\mu$
on $\Omega$. Then the diffeomorphism transformation
of $A^a_\mu$ can be
generated by $V^\mu$ via a Lie derivative
${\cal L}_VA_\mu^a = D_\mu(A^a_\nu V^\nu) + V^\nu F^a_{\nu\mu}\;,$
so that ${\cal L}_VI_{CS}[A] = 0$,
provided the space-time manifold $\Omega$ has no boundary
or $V^\mu$ vanishes on the boundaries \cite{Ho}\cite{Chen}.
The first term in the diffeomorphism transformation of
the Chern-Simons field is the gauge transformation with the
gauge parameter $\tau^a = A^a_\nu V^\nu$,
and the second term is proportional to the flat
connection condition. Therefore,
on the constraint surface, a generator
generates simultaneously both diffeomorphism
and gauge transformation.
As a topological quantum field theory,
(\ref{Scs}) is exactly soluble \cite{W2}.

Obviously, the physical reality at low energies
requires symmetry breaking so that at least
local dynamical degrees of freedom emerge.
There are several possible
sources of topological symmetry breaking.
First, to quantize any gauge theory, gauge fixing is needed.
Doing so for (\ref{Scs}), it is unavoidable to pick up a
metric on $\Omega$. However, as long as
 topological invariants
such as the Wilson lines that do not
require a choice of metric are calculated,
outcomes are topologically invariant too \cite{W2}.
Other sources discussed in literature
include the functional volume element
${\cal D}A_\mu$ which
depends on the Riemannian structure
of $\Omega$ in the path
integral; and regularization and
renormalization in quantization.
On the other hand, to involve local dynamical
matter fields breaks definitely the
topological symmetry.
What we are going to do here is not to implement
the topological action (\ref{Scs}) with
other dynamical fields, but to give a (same) mass
to two of the three components
of the Chern-Simons field $A^a_\mu$ $(a=0,1,2)$,
and to keep the third massless.
For convenience, we rewrite the massive components as
a complex vector field
$A_\mu^1+iA^2_\mu = gB_\mu $ with
$g^2=4\pi\alpha$,
and set the massless $A^0_\mu=ga_\mu.$
Then we arrive at an action
\begin{equation}
I = \int_\Omega\left(
\frac{1}{2}\epsilon_{\mu\nu\lambda}
B^*_\mu(\partial_\nu - iga_\nu)B_\lambda
+MB^*_\mu B^\mu
+ \frac{1}{2}\epsilon_{\mu\nu\lambda}
a_\mu\partial_\nu a_\lambda\right)\;.
\label{S}
\end{equation}
Involving a spacetime metric \cite{note3} when $M \neq 0$,
(\ref{S}) describes a topological symmetry
broken phase of the $(2+1)d$ gravity.
Moreover, instead of $ISO(2,1)$, (\ref{S}) is now a
$U(1)$ theory, with
gauge transformations
$a_\mu \rightarrow a_\mu - \partial_\mu \epsilon$
 and $ B_\mu\rightarrow e^{i\epsilon}B_\mu$.
(\ref{S}) is actually a theory of a
charged massive vector $B_\mu$ coupled to
a gauge field $a_\mu$ who's dynamics is governed
by a Chern-Simons term. This theory was used
in a different context (anyon)
\cite{CI1}\cite{CI2}. We should point out
that, besides carrying a spacetime metric,
a nonvanishing mass
develops local dynamical degrees of freedom
for $B_\mu$ field. This will be explicitly seen
below. Naturally, the mass $M$ serves
as the order parameter for the symmetries.

Now we attach the system
 (\ref{S}) to a heat bath with temperature $T$.
As is known, finite temperature
behavior of any theory
is specified by the partition function
$Z={\rm Tr}e^{-\beta H};$
and the thermal expectations of physical observables
$<{\cal O}> ~= \frac{1}{Z}
{\rm Tr}[{\cal O}e^{-\beta H}],$
where $\beta = 1/T$ is the inverse
temperature ($k_B = 1$) \cite{K}.
The functional integral representation of
partition function for the system described by
(\ref{S}) is
\begin{equation}
Z = \int{\cal D}a_\mu
{\cal D}B^*_\nu
{\cal D}B_\lambda
{\cal D}\bar{c}
{\cal D}c
{\rm exp}\left( -\int_0^\beta d\tau\int d^2x
{\cal L}\right)\;,
\label{Z}
\end{equation}
with the Euclidean Lagrangian
\begin{equation}
{\cal L}=
-\frac{i}{2}\epsilon_{\mu\nu\lambda}
B^*_\mu(\partial_\nu - iga_\nu)B_\lambda
+MB^*_\mu B_\mu
-\frac{i}{2}\epsilon_{\mu\nu\lambda}
a_\mu\partial_\nu a_\lambda
+\frac{1}{2\rho}(\partial_\mu a_\mu)^2
+ (\partial_\mu\bar{c})(\partial_\mu c)\;.
\label{Se}
\end{equation}
Above $\rho$ parametrizes
the covariant gauge fixing for
$U(1)$ symmetry, and the
ghost field $c$ does not interact
with other fields but only serves to
cancel the non-physical (in fact all)
degrees of freedom of $a_\mu$.
Besides, all the fields are subject to the
periodic boundary condition such as
$a_\mu(\beta, {\bf x})=a_\mu(0, {\bf x})$.

Switching off the Chern-Simons
interaction ($g \rightarrow 0$), we consider the
free theory first.
With Fourier transformation, it is readily to obtain
the free energy density
for the free theory in the momentum space
\begin{equation}
{\cal F}_0 = -\frac{{\rm ln}Z_0}{\beta V}
= \frac{1}{\beta V}\sum_n\sum_{\bf p}
{\rm ln}\left(\frac{\beta M^2}{\rho}\right)
 + 2T\int\frac{d^2p}{(2\pi)^2}\left[\beta\omega
+ ln(1-e^{-\beta\omega})\right]\;,
\label{Z0}
\end{equation}
where $\omega = \sqrt{{\bf p}^2+M^2}$.
One sees now that the massive
vector $B_\mu$ obeys the Bose-Einstein distribution.
The factor ``$2$'' in the second term in (\ref{Z0}) indicates
two degrees of freedom,
carried by the complex field $B_\mu$, in the thermal equilibrium.
And that the contribution from the gauge field $a_\mu$
is just cancelled by that from the ghost,
except a gauge parameter $\rho$ dependent term which
contributes only the zero-point energy. This verifies
Chern-Simons field carries no local dynamical degree of
freedom.

Then, consider a perturbation
in dimensionless coupling $g$, around the
free theory.
As the Chern-Simons coupling $g$, receiving
no non-trivial renormalization, doesn't run \cite{CH}\cite{H},
it serves well as a controlling parameter
in a perturbation expansion.
{}From (\ref{Z}), 
it is easy to work out the finite temperature
Feynman rules:
\begin{equation}
G^0_{\mu\nu} = \frac{\epsilon_{\mu\nu\lambda}p_\lambda
+\delta_{\mu\nu}M+p_\mu p_\nu/M}{p^2+M^2}\;,~~
D^0_{\mu\nu}(p)
= \frac{\epsilon_{\mu\nu\lambda}p_\lambda}{p^2}\;,
 ~~{\rm and}~~
\Gamma^0_{\mu\nu\lambda}=g\epsilon_{\mu\nu\lambda}
\label{R}
\end{equation}
for the vector and Chern-Simons propagators
(in the Landau gauge) and vertex, respectively.
Due to the periodic boundary condition for
$B_\mu$ and $a_\mu$ fields, the third component of
momentum, the frequency, takes discrete values, $p_3 = 2\pi nT$
for integer $n$'s. Besides, each loop in a Feynman diagram
carries an integration-summation $T\sum_n\int d^2p/(2\pi)^2$
over the
internal momentum-frequency $({\bf p},2\pi Tn)$.

By the inverse two-point
function of $B_\mu$ field,
${G_{\mu\nu}}^{-1}(p) = {G^0_{\mu\nu}}^{-1}(p)
-\Sigma_{\mu\nu}(p)$ with the self-energy
$\Sigma_{\mu\nu}(p)$,
the effective mass of the vector field $B_\mu$ is defined as \cite{note2}
\begin{equation}
M(g,T)\delta_{\mu\nu} = {G_{\mu\nu}}^{-1}(p=0)\;.
\label{Mt}
\end{equation}
$M(g,T)$, as the order parameter, characterizes a
possible topological phase transition.
Calculating
$\Sigma^{(2)}_{\mu\nu}(p) = g^2T\sum_n\int\frac{d^2q}{(2\pi)^2}
\epsilon_{\mu\sigma\eta}
G^0_{\sigma\lambda}(p+q)\epsilon_{\lambda\tau\nu}D^0_{\tau\eta}(q),$
the self-energy at the second order, and using (\ref{Mt}),
we obtain the effective mass
\begin{equation}
M(g,T) =M_r+\frac{1}{6\pi}g^2
\left(M_r + 2T{\rm ln}(1 - e^{-M_r/T})\right) + {\cal O}
(g^4)\;.
\label{MT}
\end{equation}
 $M_r$ denotes
zero temperature renormalized mass. For instance,
$M_r = M-\frac{1}{3\pi^2}g^2\Lambda$
at one-loop if a regularization by a naive ultraviolet cutoff
$\Lambda$ is used. In the bracket in (\ref{MT}),
the bare mass $M$ has been replaced by
the renormalized one $M_r$, this affects
only higher orders. The first term in the bracket
comes from the radiative correction at zero
temperature. The second term is due to exchanging energy
with the heat reservoir. The low temperature limit
$T \rightarrow 0$ is trivial, as $M(g,T) \rightarrow
(1+\frac{1}{6\pi} g^2)M_r > 0$.
On the other hand, since $M(g,T)$ is a monotonically
decreasing function of temperature, as $T$ goes up,
the thermal excitation tends to drive the effective mass
to zero. Namely, there must exist a critical temperature
$T_c$ at which  $M(g, T)=0$, and a phase transition
happens. $T_c$ is readily to solve from (\ref{MT}).
We obtain 
$e^{-aM_r/T_c} = 1-e^{-M_r/T_c},$  with $ a=(1/2+3\pi/g^2).$
In a linearized form, as a good approximation
when $T_c \gg M_r$,
\begin{equation}
T_c \simeq 3(\frac{1}{2} + \frac{\pi}{g^2})M_r\;.
\end{equation}
Now we see that the Chern-Simons
interaction is responsible to the phase transition
as it should be, and a stronger interaction causes
a transition at a lower temperature, with the
renormalized mass $M_r$ fixed.

To understand the nature of the phase transition, we come
to consider the free energy. 
In the symmetry phase, due to the topological nature,
the free energy must identically vanish.
As a consistent check, this can be readily verified
within the covariant quantization
used in this work. To gauge fix
the non-Abelian Chern-Simons theory (\ref{Scs}),
as usual, one introduces ghost term
$\partial_\mu \bar{c}^aD_\mu c^a$. Doing so,
one fixes diffeomorphism too. This ghost term includes
the kinetic of ghost and Yukawa interaction
between the ghost and Chern-Simons field. And then,
order by order, the contributions to quantities like
the free energy from the ghost cancel out completely
those from the Chern-Simons field, at zero temperature
\cite{CSW} as well as in the thermal ensemble.
This is a natural consequence of lacking of dynamical
degree of freedom in a topological theory.
On the other hand, in the
symmetry broken phase, the local dynamical degrees
of freedom make a non-vanishing free energy density,
of which the lowest order has been given in (\ref{Z0}).
At the second order, calculating

\unitlength=1.00mm
\linethickness{0.4pt}
\thicklines
\begin{picture}(110.0,33.0)
\put(55.00,18.00){\makebox(0,0)[cc]{${\rm ln} Z_2 = -\frac{1}{2}$}}
\put(68.00,18.00){\circle*{2.00}}
\put(82.00,18.00){\circle*{2.00}}
\put(75.00,18.00){\circle{40.00}}
\multiput(68.0,18.0)(2.00,0.00){7}{\line(3,0){1.00}}
\end{picture}
\vspace{-0.6cm}

\noindent with the real (dashed) line
standing for the $B_\mu$ ($a_\mu$) propagator, we obtain
\begin{equation}
{\cal F}_2 = - \frac{{\rm ln}Z_2}{\beta V}
=  \frac{1}{(4\pi)^2}g^2M_rT^2\left(\frac{M_r}{T}+
2{\rm ln}(1- e^{-M_r/T})\right)^2 + {\cal O}
(g^4)\;.
\label{F2}
\end{equation}
Being positive-definite, ${\cal F}_2$
increases monotonically with $T$. Physically,
this implies the quantum thermal fluctuation
tends to decrease the
pressure $P(T) = -{\cal F}(T)$, contrary to the
classical thermal behavior of the system
shown in (\ref{Z0}).  However,
putting together (\ref{Z0}) and (\ref{F2}) and dropping
the zero-point energy, it is easy to find
$P(T=T_c) > 0$  to the
order $g^2$.
This indicates a zeroth order transition.

To conclude, it is seen a discontinuous phase
transition concerning topological and gauge symmetries
at one loop order. It seems to suggest that,
as the system approaches the critical
temperature from the broken phase, the effective
mass smoothly limits to zero, the metric and
local dynamical degrees of freedom ``softly'' vanish,
but the free energy and so the pressure suddenly disappear.
This phenomenon, never observed in the nature nor in theories
as we know, looks natural to a topological theory,
and it deserves further studies.

The author thanks L. Dolan, M. Li and Y.S. Wu
for discussions. This work was supported in part by
the  U.S. DOE under contract No. DE-FG05-85ER-40219.


\begin{thebibliography}{199}
\bibitem{note3}
A flat spacetime metric is assumed here for simplicity,
though the mass term itself makes the gravity non-``trivial''.
This shouldn't qualitatively affect
the possible phase transition.
\bibitem{note2} With only $O(2)$ symmetry in the space,
instead of $SO(2,1)$,
 $M(g,T)$ in the longitudinal and transverse
directions are unnecessary the same in general.
However, in this example we consider, it is the case.
\end{thebibliography}
\end{document}